\documentclass[12pt,preprint]{aastex}

\shorttitle{STATISTICAL STUDY OF THE RECONNECTION RATE}

\begin{document}

\title{STATISTICAL STUDY OF THE RECONNECTION RATE IN SOLAR FLARES OBSERVED WITH {\itshape YOHKOH}/SXT}

\author{Kaori Nagashima\altaffilmark{1} and 
Takaaki Yokoyama\altaffilmark{2}}

\altaffiltext{1}{Department of Astronomy, Kyoto University, Sakyo-ku, Kyoto 606-8502, Japan; kaorin@kwasan.kyoto-u.ac.jp}
\altaffiltext{2}{Department of Earth and Planetary Science, University of Tokyo, Hongo, Bunkyo-ku, Tokyo 113-0033, Japan}

\begin{abstract}
We report a statistical study of flares observed with 
the Soft X-ray Telescope (SXT) onboard {\itshape Yohkoh} 
in the year of 2000.
We measure physical parameters of 77 flares,
such as the temporal scale, the size, and 
the magnetic flux density and find that
the sizes of flares tend to be distributed more broadly
as the {\itshape GOES} class becomes weaker 
and that there is a lower limit of magnetic flux density that 
depends on the {\itshape GOES} class.
We also examine the relationship between these parameters and
find weak correlation between temporal and 
spatial scales of flares.
We estimate reconnection inflow velocity,
coronal Alfv\'en velocity, and reconnection rate 
using above observed values.
The inflow velocities are distributed from 
a few ${\rm km \ s^{-1}}$ to several
tens ${\rm km \ s^{-1}}$ and the Alfv\'en velocities
in the corona are in the range from $10^3$ to $10^4$ 
${\rm km \ s^{-1}}$.
Hence the reconnection rate is $10^{-3}$ - $10^{-2}$.
We find that the reconnection rate in a flare tends to decrease
as the {\itshape GOES} class of the flare increases.
This value is within one order of magnitude from the 
theoretical maximum value predicted by the Petschek model,
although the dependence of the reconnection rate on the 
magnetic Reynolds number tends to be stronger than that in
the Petschek model.
\end{abstract}

\keywords{Sun: corona--- Sun: flares --- Sun: magnetic fields}

\section{INTRODUCTION} \label{sec:intro}

In the energy release process in solar flares,
magnetic reconnection is generally considered to play a key role.
The reconnection rate is an important quantity,
because it puts critical restrictions on the reconnection model.
It is defined as $M_{\rm A } \equiv V_{\rm in}/V_{\rm A }$ 
in non-dimensional form,
where $V_{\rm in }$ is the velocity 
of the reconnection inflow and
$V_{\rm A }$ is the Alfv\'en velocity.
It gives the normalized value of the reconnected 
flux per unit time. In spite of its importance, 
what determines the reconnection rate 
in flares is still a question.
In the steady reconnection model of
\citet{1958IAUS....6..123S} and \citet{1957JGR....62..509P},
the reconnection rate is $M_{\rm A}=(R_{\rm m })^{-1/2}$, where 
$R_{\rm m }=(V_{\rm A } L/\eta)$
is the magnetic Reynolds number defined with the Alfv\'en velocity
and $\eta$ is the magnetic diffusivity.
In the solar corona, if the resistivity is attributed 
to the Coulomb collision \citep{1956pfig.book.....S}, 
the typical value of $R_{\rm m}$ is 
$R_{\rm m}\sim 10^{14}$, which means $M_{\rm{A}}\sim 10^{-7}$ 
and estimated timescale of flare is about one year. 
It is, of course, too slow to explain flares 
whose timescale is about $10^2-10^3$ seconds.
On the other hand, \citet{1964psf..conf..425P} pointed out that 
the previous model lacks the effect of wave and suggested 
his model with $M_{\rm A} \lesssim \pi/(8 \ln{(8  R_{\rm m})})$ .
The special feature of this model is that $M_{\rm A}$ has weak
dependence on $R_{\rm m}$. In this model
 $M_{\rm A}\lesssim 10^{-2}$ when $R_{\rm m} \sim 10^{14}$ and 
the estimated timescale is consistent with observed value
(see also \citealt{2004ApJ...610.1117N}).

The Sweet-Parker model and the Petschek model shown above are 
steady reconnection models.
The physics of nonsteady reconnection is not established clearly.
\citet{2001EP&S...53..473S} discussed the
plasmoid-induced-reconnection model.
Since flares are nonsteady phenomena,
they suggested possibility that the reconnection rate in a flare 
is determined by the nonsteady and dynamic processes, 
such as plasmoid ejections.
Another uncertainty in the model of the reconnection mechanism
is uncertainty of the magnetic Reynolds number $R_{\rm m}$.
It was suggested that finite levels of MHD turbulence can lead to 
increase of the magnetic diffusivity, i.e., 
decrease of the effective
magnetic Reynolds number (\citealt{1985PhFl...28..303M};
\citealt[\S 3.3]{1997plas.conf.....T}).

Though there are some observational case studies 
about the reconnection rate, 
statistical studies are very few so far.
\citet{1996ApJ...472..864D} examined the spatial 
and temporal scales of
reconnection events in the solar atmosphere described in the 
literature and found that the reconnection rate is on the 
order of 0.001 - 0.1.
\citet{1996ApJ...456..840T} analyzed the temperature structure 
of 1992 February 21 LDE flare in detail and
found that the reconnection rate is 0.07.
He assumed isothermal switch-off shock and used
the condition of the shock to derive the reconnection rate.
The flare on 1992 January 13 was 
analyzed by \citet{1997ApJ...478..787T} 
and the reconnection rate in this flare was estimated at 0.06.
\citet{1997PASJ...49..249O} and \citet{1998ApJ...499..934O} studied
1993 November 11 flare and 1992 October 5 flare, respectively.
They found the value of 
0.003 - 0.2 in preflare and impulsive phase,
and 0.0002 - 0.0003 in gradual phase.
\citet{2002ApJ...566..528I} analyzed the decay phase of 
an LDE flare on 1997 May 12 and estimated the reconnection rate
at 0.001-0.01 using the energy release rate
and the reconnected magnetic flux.
Using the same method as the previous one,
\citet{2002mwoc.conf..171I} analyzed 7 flares and
 2 giant arcades and
found that the reconnection rate is 0.001 - 0.1.
Moreover, analyzing the impulsive phase of three two-ribbon flares,
\citet{2005ApJ...632..1184I} obtained the value of 0.015 - 0.071. 

Recently, direct measurements of 
the reconnection inflow were reported.
\citet{2001ApJ...546L..69Y} found an evidence of
reconnection inflow and derived the reconnection rate
as 0.001 - 0.03. By the same method,
\citet{Narukage_2006} reported 6 inflow events 
and found the reconnection rate is 0.001 - 0.07.
\citet{2005ApJ...622..1251L} studied an eruptive limb event 
occurred on 2003 November 18 and reported the reconnection rate
is in the range from 0.01 to 0.23. 
Note also that they found the reconnection inflow
using the Doppler shift.

In this paper, we report a statistical study of 
flares that occurred in the year of 2000 observed with 
{\itshape Yohkoh}/SXT.
We determine several physical quantities of each flare
and try to estimate the reconnection rate in solar flares.
Our aim is to find statistical tendency of the physical parameters.
In \S \ref{sec:analysis}, we describe the data and
 the method of analysis.
In \S \ref{sec:result}, we show results.
Discussion and conclusion are given in \S \ref{sec:discuss} and 
\S \ref{sec:conclusion}, respectively.

\section{DATA ANALYSIS} \label{sec:analysis}

\subsection{Inflow Velocity}
The amount of energy released during a flare $E_{\rm flare}$ can 
be explained by the magnetic energy stored in the solar atmosphere;
\begin{equation}
E_{\rm flare}\sim E_{\rm{mag}} =\frac{B_{\rm cor}^2}{8\pi} L^3 ,
\label{eq:Eflare}
\end{equation}
where $L $ is the characteristic size of the flare and 
$B_{\rm cor}$ is the characteristic magnetic flux density in the corona.  
Since the released magnetic energy balances   
the energy flowing into the reconnection region,
we can describe the energy release rate as 
\begin{equation}
\left|\frac{dE_{\rm{mag}}}{dt} \right| \simeq 2 \times \frac{B_{\rm cor}^2}{4\pi}V_{\rm{in}}\times L^2 , \label{eq:dEdt}
\end{equation}
where $V_{\rm in}$ is the inflow velocity of the plasma.
Therefore, the time required for the energy inflow to 
supply the flare energy is estimated as
\begin{equation}
 \tau_{\rm{flare}}\sim E_{\rm{flare}}\left( 
\left|\frac{dE_{\rm{mag}}}{dt}\right|\right)^{-1} \sim \frac{L}{4 V_{\rm{in}}}
\end{equation}
and this should be the timescale of the flare.
Using this timescale, we can estimate the inflow velocity 
$V_{\rm in}$ as
\begin{equation}
V_{\rm in} \sim \frac{L}{4\tau_{\rm flare}} . \label{eq:vin}
\end{equation}
To evaluate the reconnection rate in non-dimensional form 
$M_{\rm A} \equiv V_{\rm in}/V_{\rm A}$, 
we must estimate 
the Alfv\'en velocity in the inflow region 
$V_{\rm A}=B_{\rm cor}/\sqrt{4\pi \rho}$.
Hence, if we measure the coronal density $\rho$,
the spatial scale of the flare $L$, 
the magnetic flux density in the corona $B_{\rm cor}$, and
the timescale of flares $\tau_{\rm flare}$, 
we can calculate
inflow velocity $V_{\rm in}$, 
Alfv\'en velocity $V_{\rm A}$,
and reconnection rate $M_{\rm A}$.

\subsection{How to Determine the Physical Parameters of 
the Solar Flare}
\label{sec:howto}
In this subsection we describe how to determine 
the parameters required for derivation of 
the inflow velocity $V_{\rm in}$ and
the Alfv\'en velocity $V_{\rm A}$.
To estimate the inflow velocity $V_{\rm in}$ 
by using equation (\ref{eq:vin}),
we need the spatial and temporal scales of each flare:  
$L$ and $\tau_{\rm flare}$.
First, the flare size $L$ is determined by 
soft X-ray images taken by the Soft X-ray Telescope 
(SXT; \citealt{1991SoPh..136...37T}) onboard
the {\itshape Yohkoh} satellite \citep{1991SoPh..136....1O} 
with the beryllium (Be) filter.
We define $L$ as the square root of the area 
where the intensity is more than 10\% of the maximum value 
in each frame and
it is measured at the time when {\itshape GOES} 
X-ray intensity attains its 
maximum value in each event.
The error of $L$ is considered as the difference
in definition of the size;
the minimum size is defined by the area 
where the intensity is more than 30\% 
of the maximum value in each frame 
and the maximum size is defined by the area 
where the intensity is 
more than 3\% of the maximum value.
The timescale of a flare $\tau_{\rm{flare}}$ is
defined as the period from the onset to the peak of 
{\itshape GOES} X-ray flux\footnote{http://www.ngdc.noaa.gov/stp/SOLAR/ftpsolarflares.html}. 
Since we use one-min average of {\itshape GOES} X-ray flux data,
the error of this timescale is considered as one minute.
In equations (\ref{eq:Eflare}) and (\ref{eq:dEdt}), the spatial scale of the flare $L$ should be 
the length of the current sheet. Although we cannot measure it or 
estimate the ratio of the current sheet to the global size accurately, 
we can consider the length of the current sheet is the same order as 
the size of system since the flare size $L$ is measured in the impulsive phase.

To estimate the Alfv\'en velocity in the 
inflow region $V_{\rm A}$,
we need coronal density $\rho$ and coronal magnetic flux 
density $B_{\rm cor}$.
Owing to difficulty in measuring the coronal density $\rho$,
we assume it $1.67 \times 10^{-15}\  \rm{ g \ cm^{-3}}$ 
that is the typical value in the active region.
The coronal magnetic flux density $B_{\rm cor}$ is also 
difficult to measure directly.
Therefore, we try to evaluate it in two ways.
In the first method (method 1),
we use the photospheric magnetograms from the
Michelson Doppler Imager (MDI; \citealt{1995SoPh..162..129S})
onboard {\itshape the Solar and Heliospheric Observatory} 
({\itshape SOHO}; \citealt{1995SoPh..162....1D}).
We define the representative magnetic flux density 
for each flare as follows:
First, using soft X-ray images taken by SXT with the Be filter,
we determine the area in which the intensity is larger than
a threshold value. 
Second, we measure the magnetic flux
in the area and obtain the average of 
unsigned magnetic flux density.
Changing this threshold value from 10\% to 50\%
of the maximum soft X-ray intensity of each image,
magnetic flux densities are obtained as a function of the 
threshold value of the soft X-ray intensity.
Finally, we define representative 
magnetic flux density $B_{\rm ph}$ 
as the midpoint of the maximum and the minimum of 
the obtained flux density in this range and
consider the extent from the maximum to the minimum 
as the error range of $B_{\rm ph}$.

\begin{figure}[htpb]
\plottwo{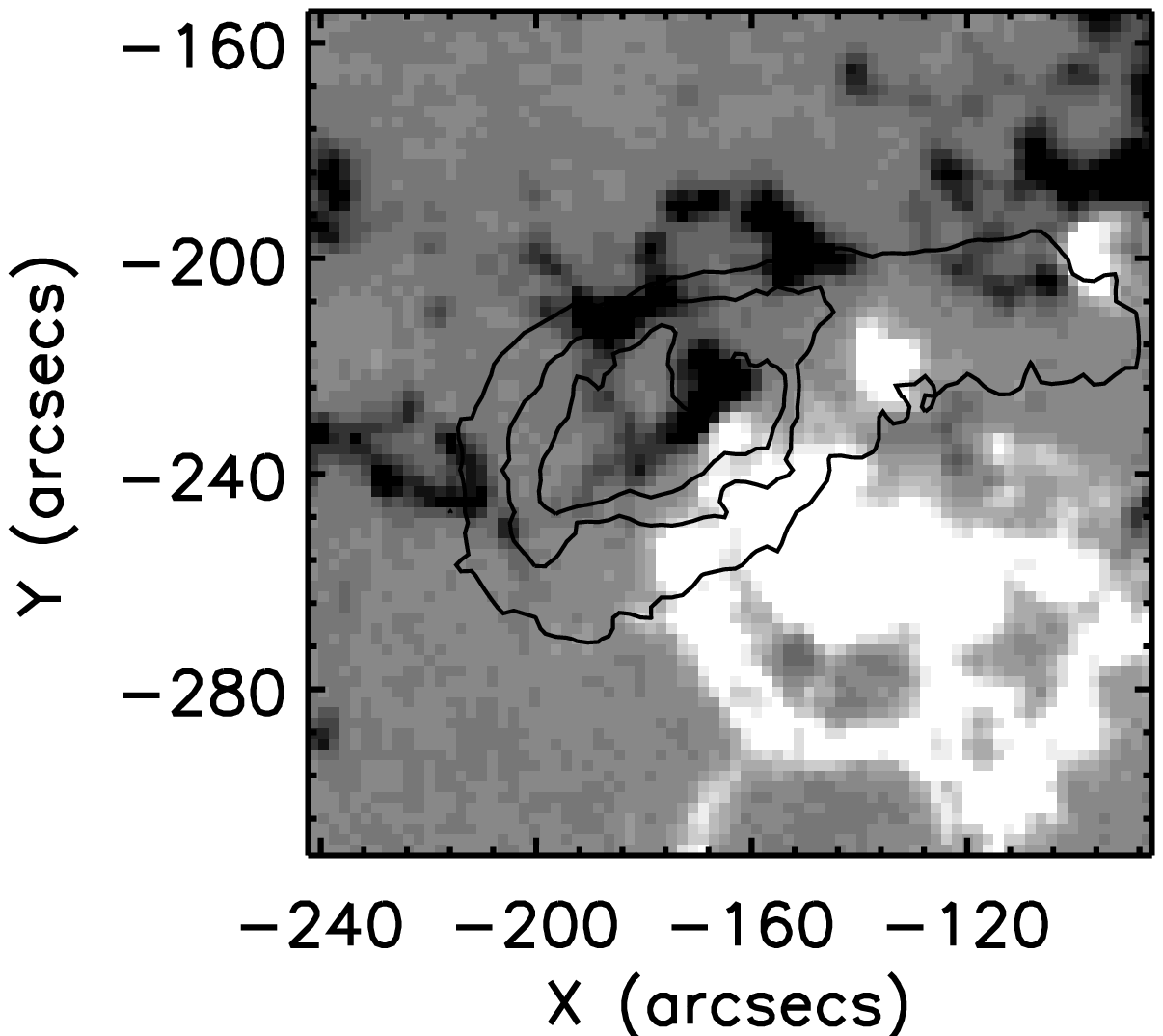}{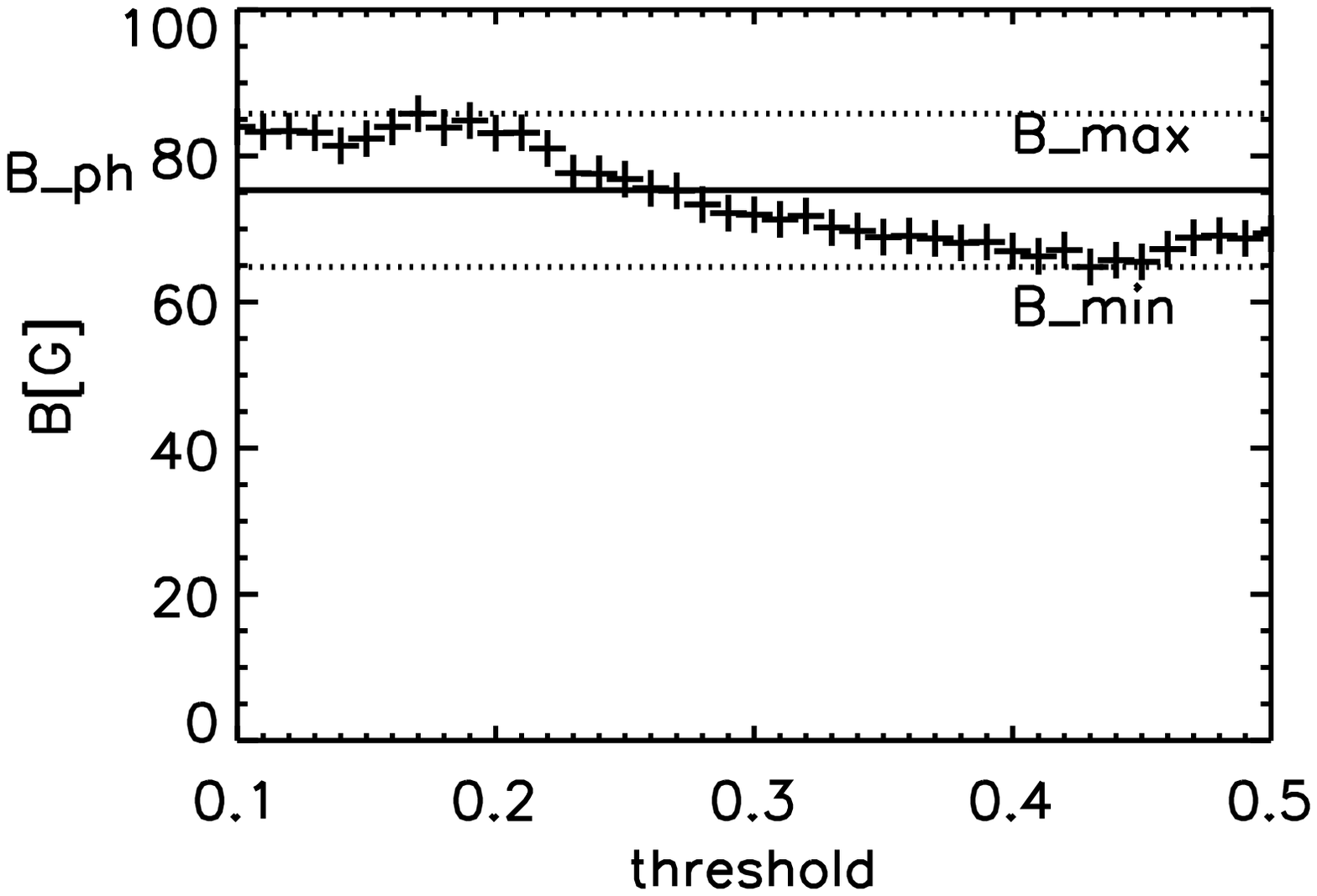}
\caption{
Left panel: Soft X-ray contours overlaid on the 
{\itshape SOHO}/MDI magnetogram of the M3.9 flare 
on 2000 January 18.
The contours show {\itshape Yohkoh}/SXT image; 10, 30, 
and 50\% of the maximum soft X-ray intensity.
White and black indicate positive and negative polarities in the 
magnetogram.
Right panel: Photospheric flux density plotted against 
threshold value (see text). 
The dotted lines indicate the maximum and minimum of the 
flux density and the solid line corresponds to the midpoint of 
the maximum and the minimum.
We define this midpoint as the representative 
photospheric flux density of this event and 
the error range as the extent from the minimum to the
maximum.}
\label{fig:Bmaxmin}
\end{figure}

In Figure \ref{fig:Bmaxmin} we show the result 
of the {\itshape GOES} M3.9 class flare on January 18
as an example.
The left panel of Figure \ref{fig:Bmaxmin} 
illustrates the photospheric magnetogram
of the flare region.
White and black indicate positive and negative polarities in
the magnetogram.
The overlaid contours correspond to 10, 30, and 50\% of the
maximum soft X-ray intensity. 
The right panel of Figure \ref{fig:Bmaxmin} 
shows photospheric magnetic flux density
as a function of the threshold value of the soft X-ray intensity.
The dotted lines indicate the maximum and minimum
flux density and the solid line corresponds to
the midpoint value, which is defined as 
the representative photospheric flux density of this event.

In order to obtain the coronal magnetic flux density,
we assume 
\begin{equation}
B_{\rm cor} = \alpha_{B} B_{\rm ph} \label{eq:BBratio}
\end{equation}
and
\begin{equation}
\alpha_{B} = {\rm const.} =0.3 \label{eq:alpha}
\end{equation}
 among all flares.
The validity of this assumption will be discussed 
in \S \ref{sec:alphadis}.

The second method (method 2) to obtain the coronal flux density
is as follows.
If all the magnetic energy stored in the corona 
before the reconnection process 
turns into the thermal energy in the reconnected loop,
we obtain 
\begin{equation}
p_{\rm loop} \sim \frac{B_{\rm eq}^2}{8 \pi},
\label{eq:PressureBalance} 
\end{equation}
where $p_{\rm loop}$ is the plasma pressure in the flare loop and
$B_{\rm{eq}}$ is the magnetic flux density in the corona.
To distinguish this magnetic flux density 
from that derived by method 1, we use the subscript ``eq''.
The value of $p_{\rm loop}$ is estimated from the 
color temperature $T$ and the emission measure $EM$ 
calculated from the
{\itshape GOES} X-ray data \citep{1994SoPh..154..275G}, that is,
\begin{equation}
p_{\rm loop} =n_{\rm{loop}}k_{\rm{B}}T,
\end{equation}
where
\begin{equation}
n_{\rm{loop}} \sim 2 \sqrt{\frac{EM}{L^3}}, \label{eq:nEMV}
\end{equation}
$n_{\rm{loop}}$ is the number density in the flare loop, and
$k_{\rm{B}}$ is the Boltzmann constant.
Since the {\itshape GOES} X-ray flux data have no spatial resolution,
we use the flare size measured in the SXT image
as the size $L$ in equation (\ref{eq:nEMV}).
Hence, we can estimate $B_{\rm eq}$ and assume 
$B_{\rm cor} \sim B_{\rm eq}$.
The error of this flux density $B_{\rm eq}$ 
results from the error in deriving
temperature and emission measure from
the {\itshape GOES} X-ray data.
On the basis of the description in \citet{1994SoPh..154..275G},
estimated uncertainty in the derived temperature depends on 
the temperature and is calculated at 13\% around 10 MK.
We calculate this uncertainty as a function of temperature and
then obtain the propagated error of $B_{\rm eq}$.

\subsection{Event Selection}
As we describe in the previous subsection,
we use SXT images, MDI magnetograms, and {\itshape GOES} 
X-ray flux data.
We survey flares that occurred in 2000 
under the following criteria:
(1) {\itshape GOES} class of the flare is above C6.0,
(2) the flare site is included in the field of view of 
the partial-frame SXT image with the Be filter 
taken at the peak of the 
{\itshape GOES} soft X-ray flux, and 
(3) the distance from the disk center to the flare region is 
within 800 arcsecs 
to avoid the line-of-sight effect in
estimation of the magnetic flux density.
From the flare list obtained by {\itshape GOES} data,
482 flares meet the above criterion (1) and
77 of those flares satisfy all the criteria.

\section{RESULT} \label{sec:result}
\begin{figure}[htpb]
\plotone{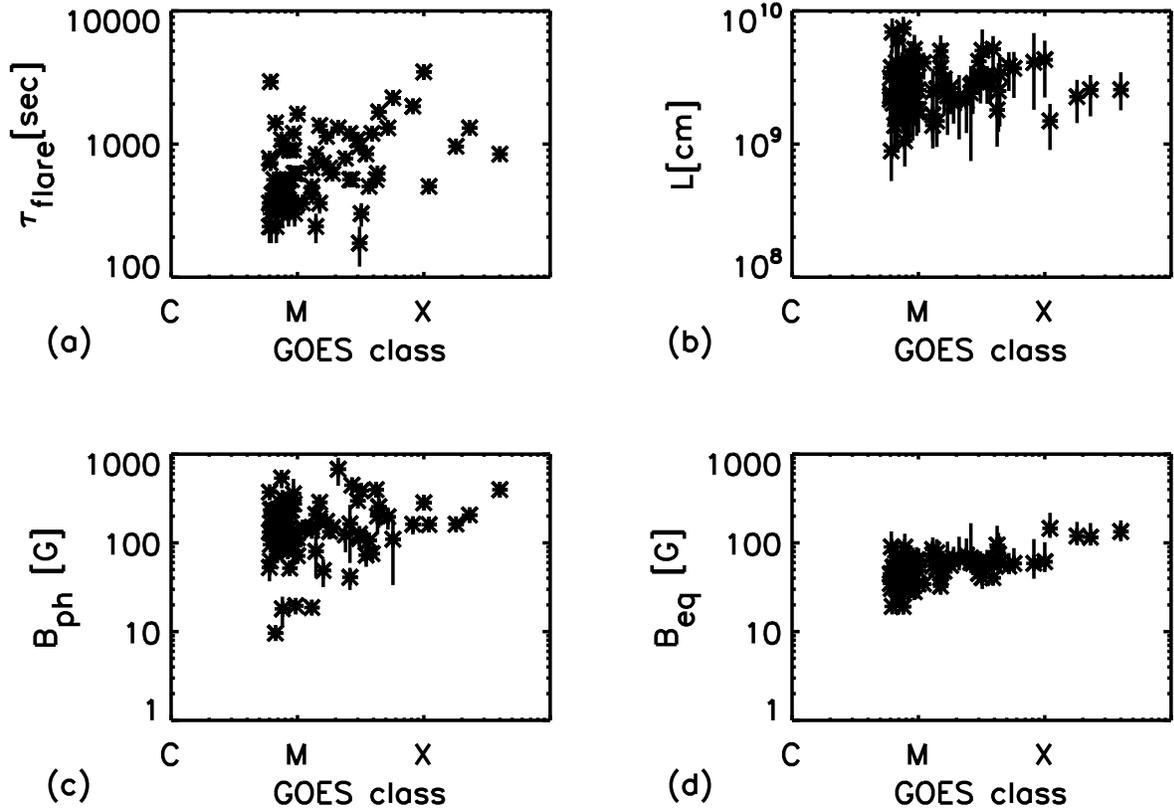} 
   \caption{Physical parameters of each flare plotted against
the {\itshape GOES} class.
(a) Time scale $\tau_{\rm flare}$. 
(b) Size $L$.
(c) Photospheric magnetic flux density $B_{\rm ph}$ 
 obtained by method 1.
(d) Coronal flux density $B_{\rm{eq}}$ obtained by method 2.
}
\label{fig:cl_para}
\end{figure}

Using the method described in the previous section,
we have analyzed 77 flares occurred in the year of 2000.
First, we examine dependence of physical parameters of these 
flares on {\itshape GOES} class.
In Figure \ref{fig:cl_para} we plot the parameters of flares
as a function of {\itshape GOES} class.
Temporal scale $\tau_{\rm flare}$, spatial scale $L$,
photospheric field $B_{\rm ph}$ obtained by method 1, and
coronal field $B_{\rm eq}$ obtained by method 2 are 
shown in Figure \ref{fig:cl_para}a, \ref{fig:cl_para}b,
 \ref{fig:cl_para}c, and \ref{fig:cl_para}d, respectively.
We find from these plots: 
(1) There is weak correlation between the timescale 
$\tau_{\rm flare}$
and the {\itshape GOES} peak flux (Panel a).  
This tendency is consistent with 
Figure 3 of \citet{2002A&A...382.1070V} and 
Figure 2 of \citet{2003A&A...400..779K}.  
(2) The characteristic size of flares $L$ 
shows a larger scatter when the {\itshape GOES} peak 
flux is smaller (Panel b). 
\citet{1998ApJ...504.1051G} reported the relation 
between the {\itshape GOES} peak flux and the loop length
of flares in their Figure \ref{fig:clMa}.
Comparing our results with theirs, 
the value of the size is in the same range,
although clear correlation between these 
parameters cannot be found out in their results.
(3) There is a threshold value in the magnetic flux density 
($B_{\rm ph}$ and $B_{\rm eq}$) that increases as 
the {\itshape GOES} peak flux increases (Panels c and d).  
For example, the X-class flares occurred 
only when $B_{\rm ph} $ was larger than $\approx$ 100 Gauss.

\begin{figure}
\plotone{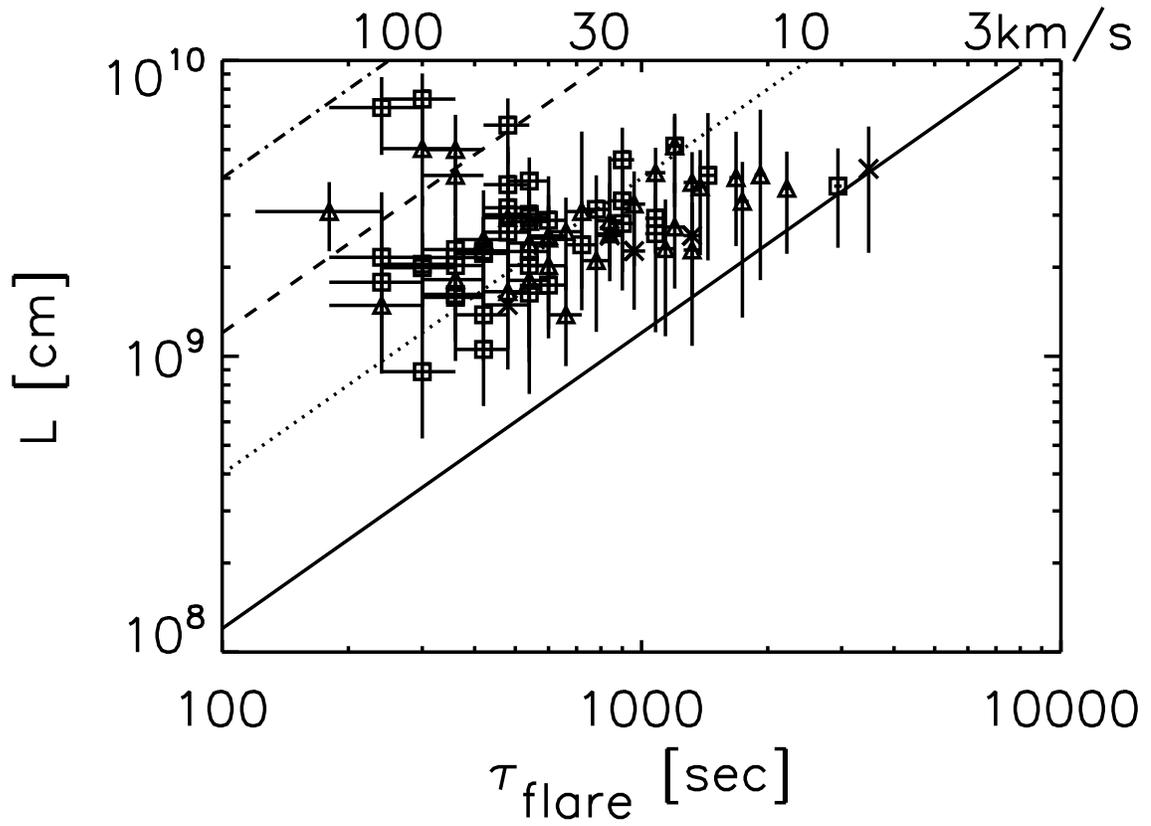}
\caption{Spatial scale of each event plotted against timescale.
Crosses, triangles, and squares indicate X-class,
M-class, and C-class, respectively.
The solid line, dotted line, dashed line, and dash-dotted 
line indicate 3, 10, 30, and 100 ${\rm km \ s^{-1}}$ 
of the inflow velocity $V_{\rm in}$, respectively.}
\label{fig:m1_tL}
\end{figure}

\begin{figure}
\plottwo{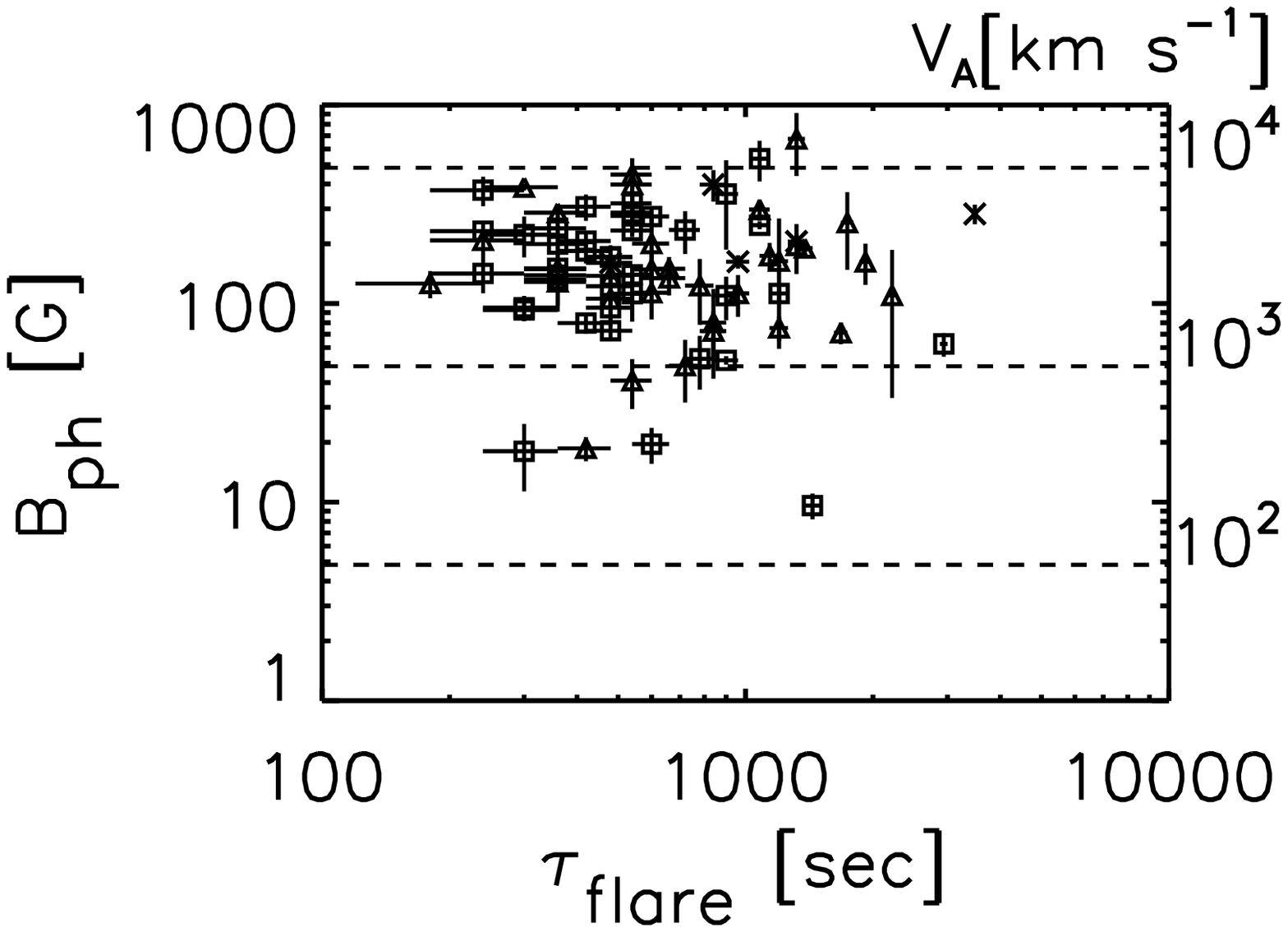}{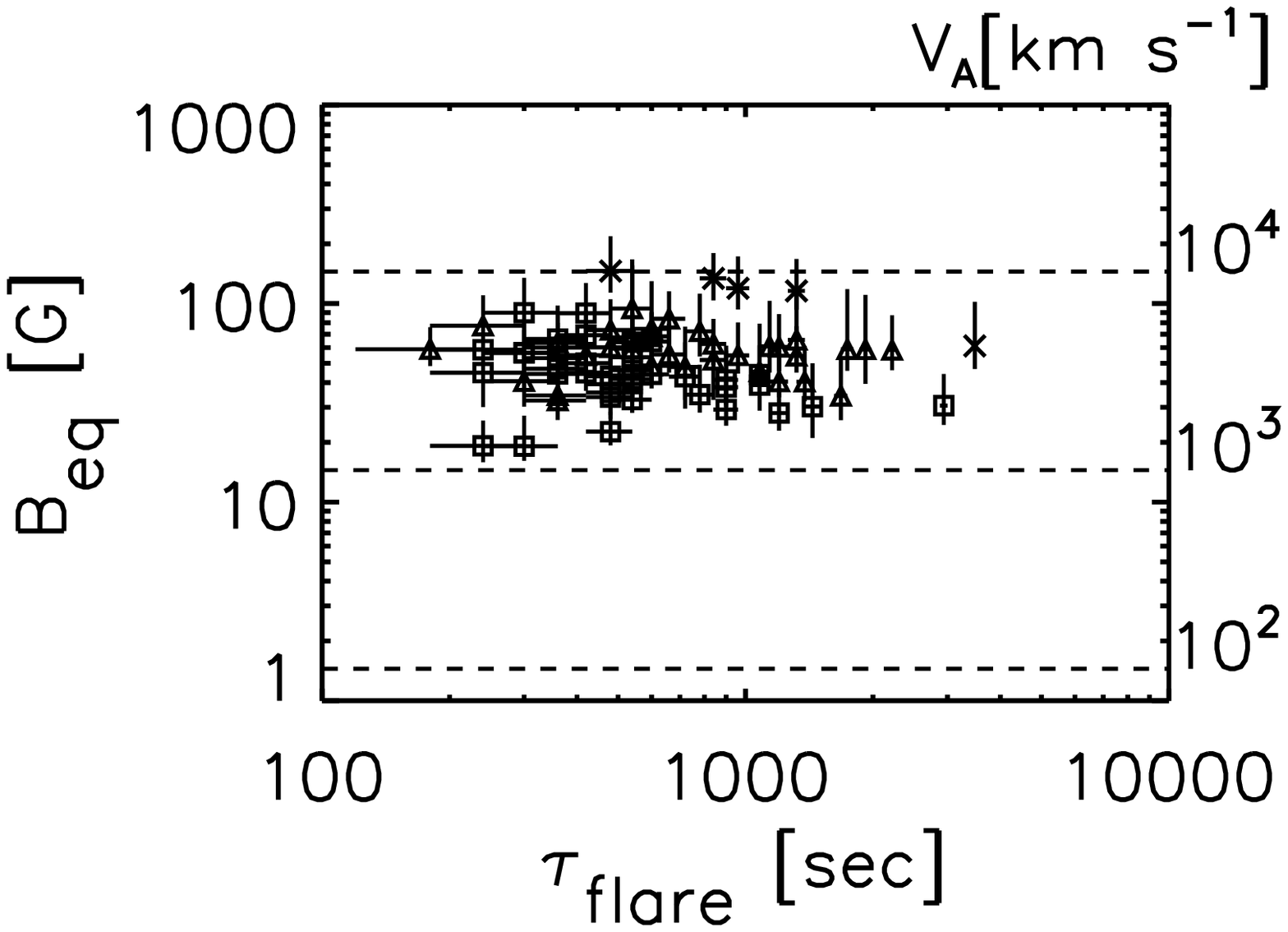}
\caption{Left panel: Photospheric flux density 
of each event $B_{\rm ph}$, obtained by method 1, 
plotted against timescale of the flares
$\tau_{\rm{flare}}$. 
The dashed lines indicate the corresponding 
coronal Alfv\'en velocity calculated on the assumption that
the coronal density 
$\rho=1.67 \times 10^{-15}\  {\rm g\  cm^{-3}}$ and
$\alpha_B \equiv B_{\rm cor}/B_{\rm ph}=0.3$.
Right panel: Coronal flux density
$B_{\rm{eq}}$, obtained by method 2, 
plotted against the timescale $\tau_{\rm{flare}}$.
The dashed lines indicate the corresponding 
coronal Alfv\'en velocity.
Meaning of the symbols in these figures 
is the same as Figure \ref{fig:m1_tL}.}
\label{fig:tB}
\end{figure}

\begin{figure}[htpb]
\plottwo{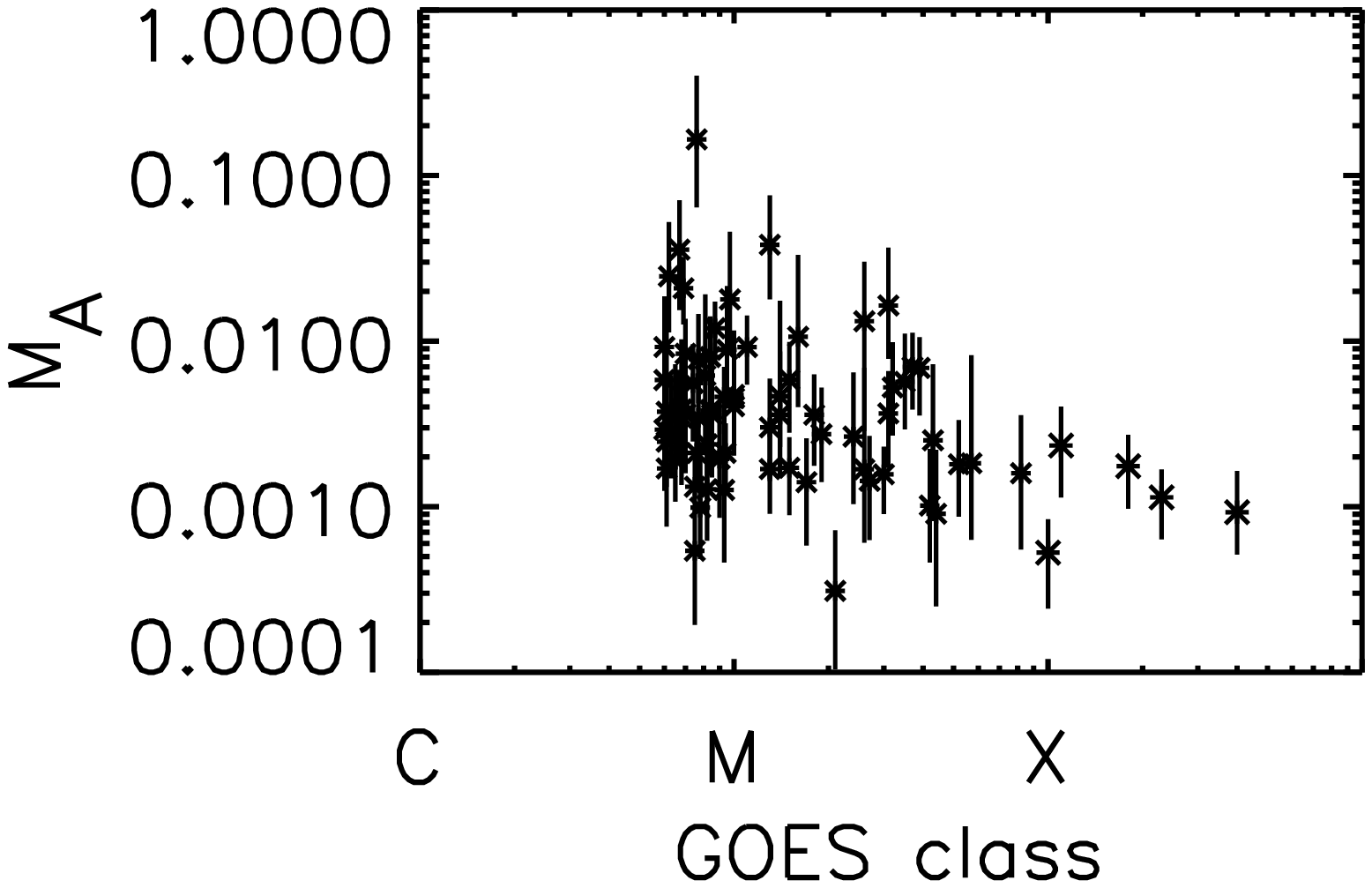}{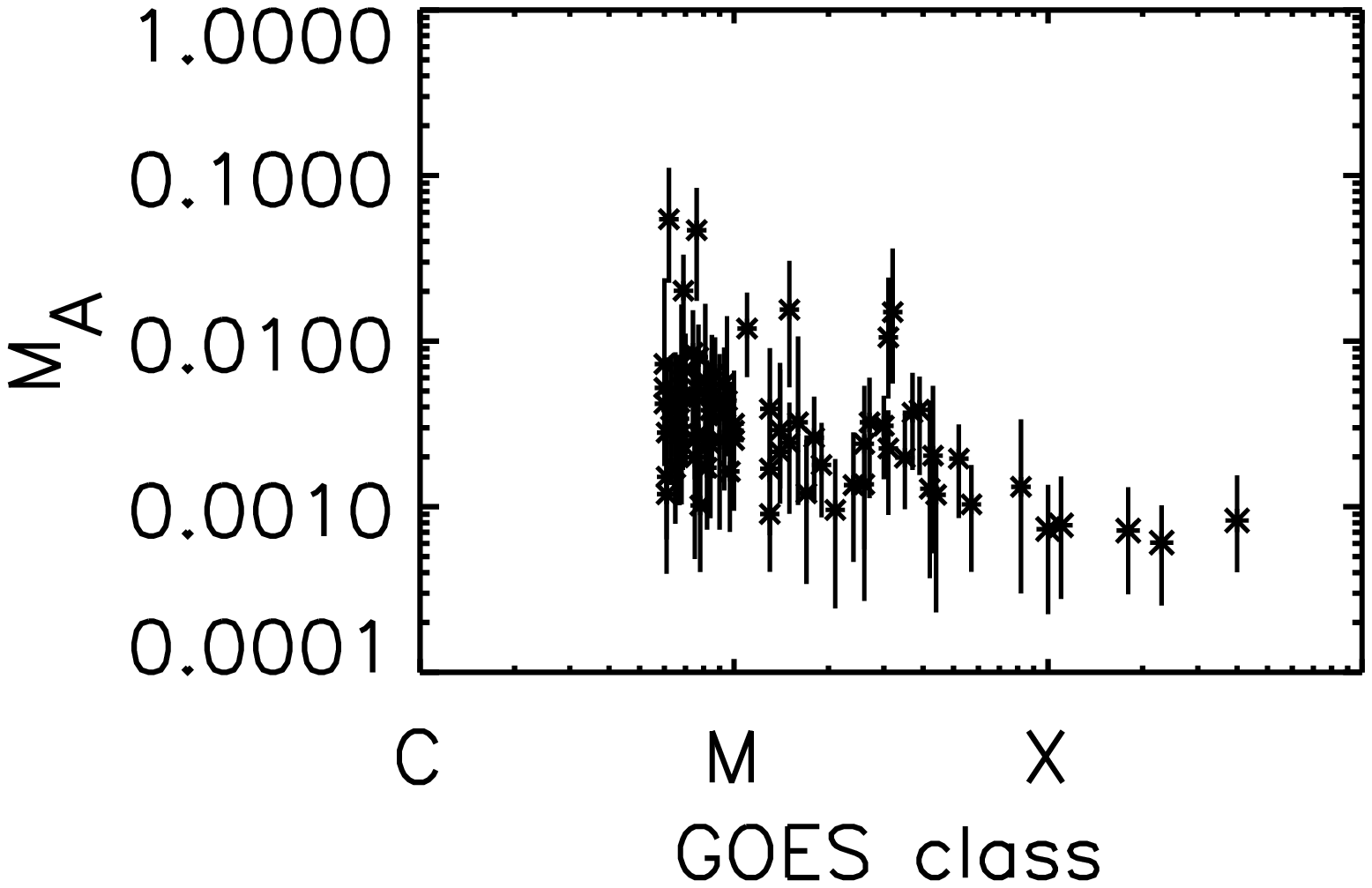}
\caption{Reconnection rate $M_A $ 
plotted against the {\itshape GOES} class of each flare.
The left panel shows the reconnection rate obtained by method 1.
We assume $\alpha_B \equiv B_{\rm cor}/B_{\rm ph} = 0.3$
when we calculate $M_A$.
The right panel shows the reconnection rate obtained by method 2. }
\label{fig:clMa} 
\end{figure}

Second, we examine the relationship between 
the physical parameters of flares.
Figure \ref{fig:m1_tL} shows the spatial scale $L$ plotted 
against the timescale $\tau_{\rm flare}$
and tells that the spatial scale $L$ tends to be 
larger with increasing timescale.
This result is consistent with the result shown in 
Figure 1c of \citet{1998ApJ...504.1051G}, 
although the tendency is not so clear in their figure 
due to a small number of their analyzed events.
The lines in Figure \ref{fig:m1_tL} indicate the inflow velocity 
$V_{\rm in}$ given by equation (\ref{eq:vin}).
The inflow velocities are found to be distributed from
a few ${\rm km \ s^{-1}}$ to several tens ${\rm km \ s^{-1}}$.
These inflow velocities are comparable to those in the previous studies.
Detailed discussion is given in \S \ref{sec:comparison}.
The left panel of Figure \ref{fig:tB} 
shows the photospheric flux density
 $B_{\rm ph}$ obtained by  method 1 
plotted against the timescale $\tau_{\rm flare}$,
and in the right panel of Figure \ref{fig:tB} the 
coronal flux density $B_{\rm eq}$ obtained by method 2
is plotted against the timescale $\tau_{\rm flare}$.
Assuming the ratio of coronal flux density to photospheric 
flux density $\alpha_B \equiv B_{\rm cor}/B_{\rm ph}$ is 
0.3 and the coronal density $\rho$ is 
$1.67 \times 10^{-15}\  \rm{g \ cm^{-3}}$, the values of magnetic 
flux density in the left panel of Figure \ref{fig:tB} 
correspond to $10^3$ - $10^4$ ${\rm km \ s^{-1}}$ of the 
Alfv\'en velocity in the inflow region $V_{\rm A}$.
The values of $B_{\rm eq}$ in the right panel of 
Figure \ref{fig:tB} also 
correspond to the same range of $V_{\rm A}$.

Using these results of $V_{\rm in}$ and $V_{\rm A}$,
we examine the reconnection rate $M_{\rm A}$.
In the left panel of Figure \ref{fig:clMa},
the reconnection rate obtained by method 1 is plotted against
the {\itshape GOES} class.
The values of reconnection rate 
are distributed around $10^{-3}$ - $10^{-2}$.
Note again that we assume $\alpha_B \equiv B_{\rm cor}/B_{\rm ph} = 0.3$
in this plot.
By changing $\alpha_B$, 
the data points will shift proportionally to the value.
The right panel of Figure \ref{fig:clMa} 
shows the reconnection rate calculated 
by using the magnetic flux density $B_{\rm eq}$ obtained 
by method 2 against the {\itshape GOES} class of each flare.
The reconnection rates are also distributed around 
$10^{-3}$ - $10^{-2}$.
Both panels of Figure \ref{fig:clMa} 
show that the reconnection rate tends to diminish
as the {\itshape GOES} class of the flare increases.

As mentioned in \S \ref{sec:howto}, 
we estimate the coronal magnetic flux density in two ways. 
In Figures \ref{fig:cl_para}c and \ref{fig:cl_para}d, 
although the flux density obtained by method 1 (Figure \ref{fig:cl_para}c) 
appears more broadly scattered than that obtained 
by method 2 (Figure \ref{fig:cl_para}d), 
the values of flux density show the similar distribution.
Moreover, the magnetic flux density 
obtained by these two methods has a similar
distribution as a function of $\tau_{\rm flare}$ 
(Figure \ref{fig:tB}).
Two panels of Figure \ref{fig:clMa} show that the values 
of the reconnection rate obtained by 
methods 1 and 2 are the same range and 
display the same dependence on the {\itshape GOES} class.
Hence, we conclude that the results 
by these two methods are consistent with each other.

\section{DISCUSSION} \label{sec:discuss}

\subsection{Ratio of Coronal Flux Density to Photospheric Flux Density} 
\label{sec:alphadis}

\begin{figure}[htpb]
\plotone{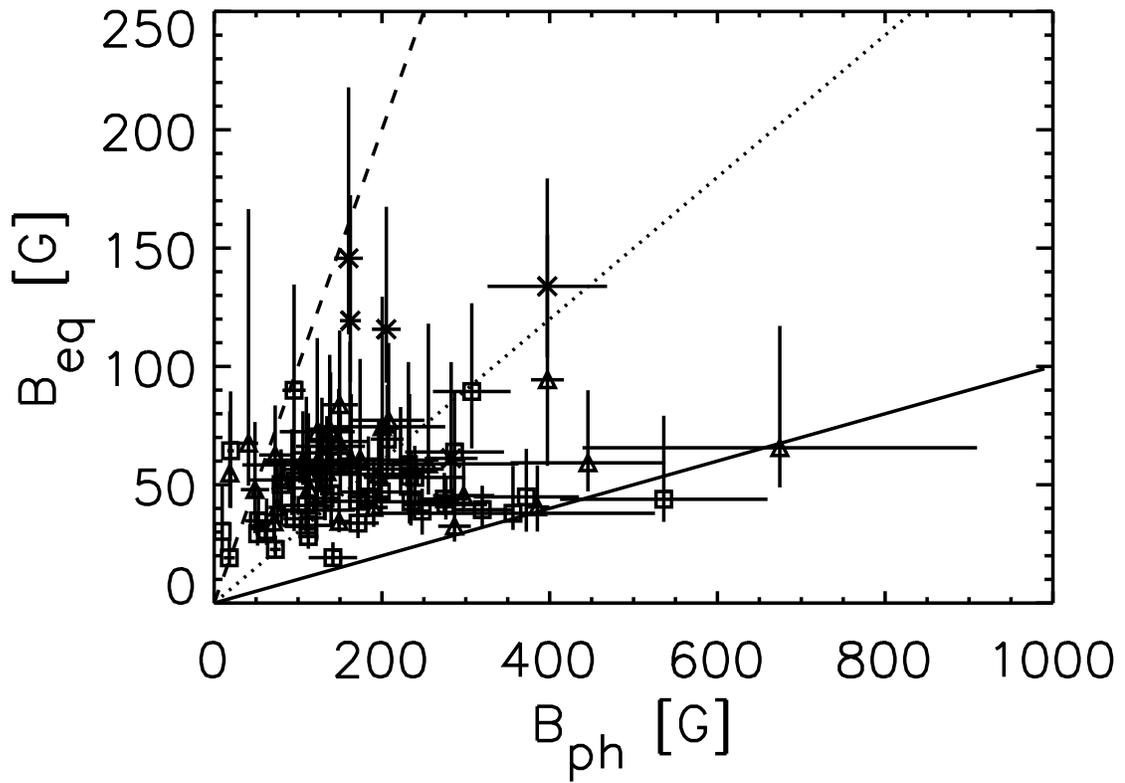}
\caption{Coronal magnetic flux density $B_{\rm eq}$  
plotted against photospheric flux density $B_{\rm ph}$.
The solid line, the dotted line, and the dashed line correspond to 
 0.1, 0.3, and 1 of $B_{\rm eq}/B_{\rm ph}$, respectively.
Meaning of the symbols is the same as Figure \ref{fig:m1_tL}.}
\label{fig:BphBeq} 
\end{figure}

We must examine whether the assumption that
the ratio of coronal flux density to photospheric flux density 
$\alpha_B \equiv B_{{\rm cor}}/B_{{\rm ph}}$ equals $0.3$ is
appropriate one.
In Figure \ref{fig:BphBeq}, we compare 
the photospheric flux density $B_{\rm ph}$
obtained by method 1 with 
the equipartitional flux density $B_{\rm eq}$ obtained by method 2.
The lines in the figure correspond to 
0.1, 0.3, and 1 of $B_{\rm eq} / B_{\rm ph}$.
This figure shows the ratios are 
distributed around 0.3, and 
if we consider  $B_{\rm eq}$ as the coronal flux density,
our assumption $\alpha_B = 0.3 $ is not inappropriate.

\begin{figure}[htpb]
\plottwo{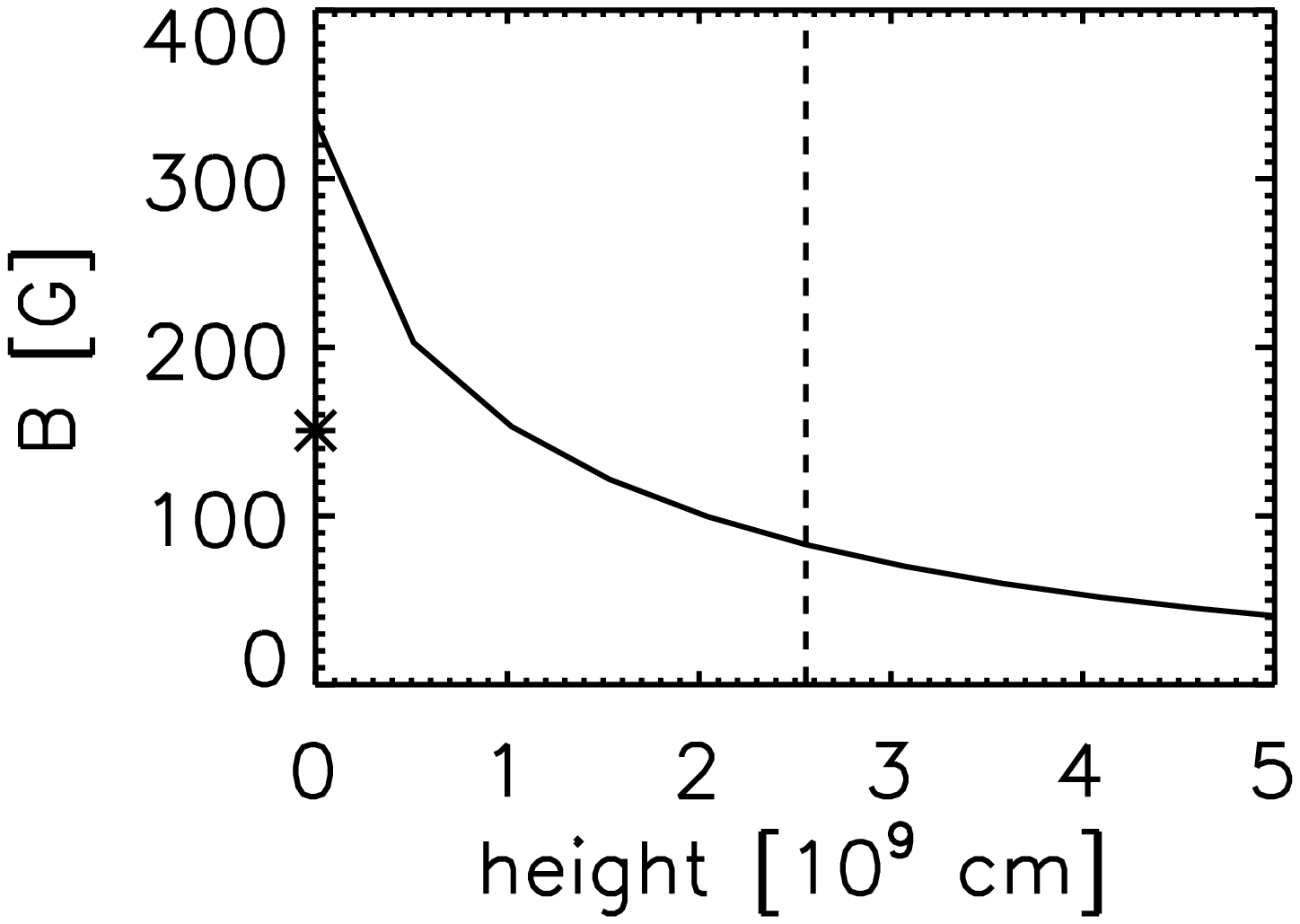}{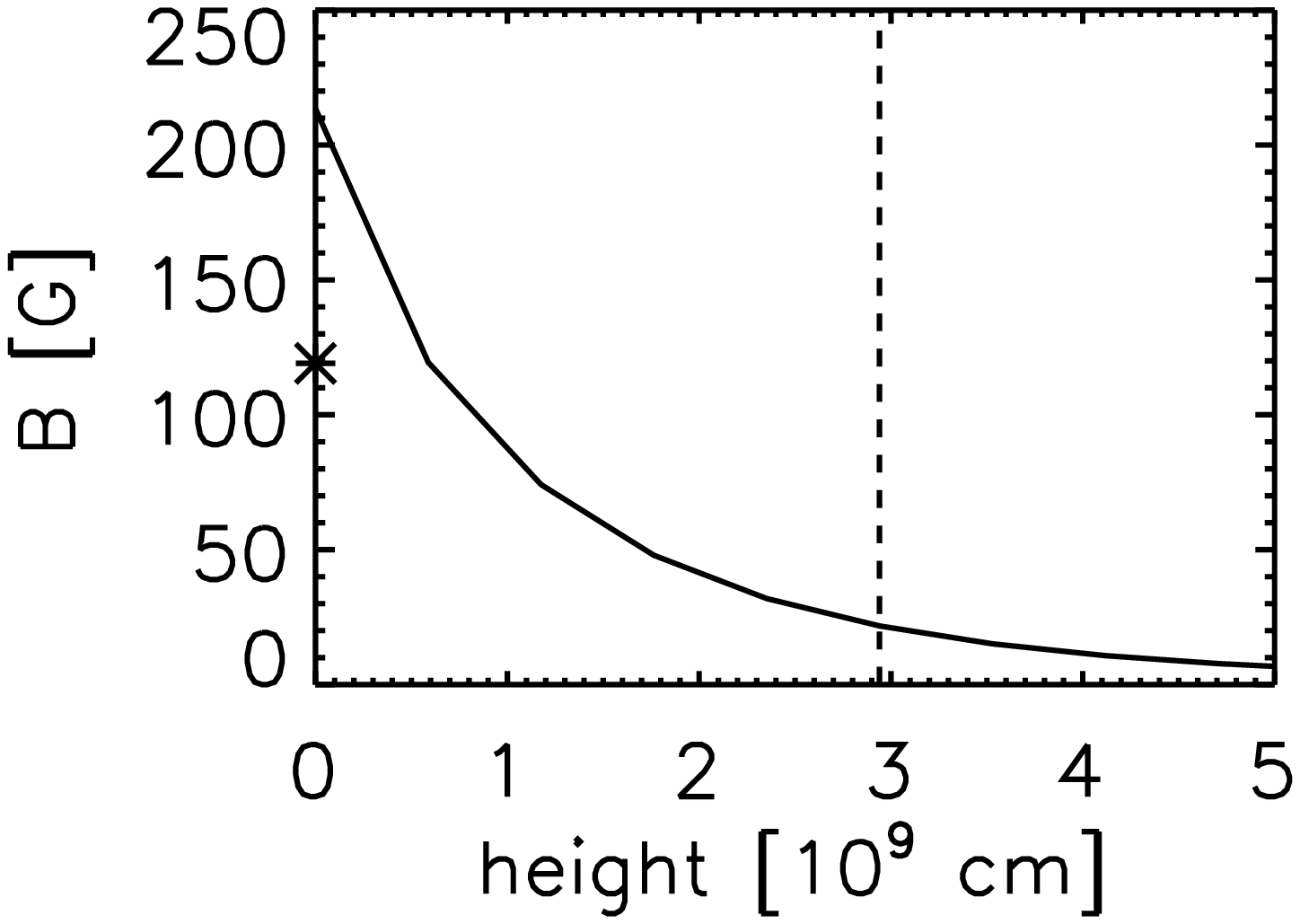}
\caption{Spatial average of potential-field flux density
plotted as a function of height.
The left panel shows an example of 2000 November 24 X2.3 flare
and the other example of 2000 July 14 M3.7 flare is shown in 
the right panel. 
The dashed line indicates the size of each flare and 
the asterisk represents the line-of-sight component of the 
photospheric flux density. }
\label{fig:potential}  
\end{figure}

\citet{1996ApJ...472..864D} estimated coronal flux density from
data of photospheric flux density
described in the scientific literature.
He calculated typical magnetic flux density as a function of height
as photospheric magnetogram forms the base.
The formula for his analyzed active region observed 
on 1974 January 16 is
\begin{equation}
B(h)= \left[ 45 \exp\left(\frac{ -h }{3.5\times 10^8 {\rm cm }}\right) 
+ 50 \exp\left(\frac{ -h }{3.0 \times 10^9 {\rm cm }}\right)
\right] \  {\rm G}, \label{eq:magDere}
\end{equation}
where $h$ is height above the photosphere.
He applied this formula to calculate the coronal flux density
of all his analyzed events in active regions.
Using the typical spatial scale of flare loops in our analysis
$L\sim 10^{9}\  {\rm cm}$ as the height $h$,
we obtain the coronal field of 38 G by equation (\ref{eq:magDere});
we can also calculate the photospheric flux density of his model 
at 95 G by taking $h=0$ in equation (\ref{eq:magDere}).
Therefore the ratio of the coronal flux density to 
the photospheric flux density is roughly estimated at 0.4.
This formula probably underestimates the magnetic flux density,
because the photospheric flux density 
analyzed in our paper is in the range 
from 100 G to 700 G.
Hence, we also did a calculation of
the coronal potential field for some particular events
for more investigations.
The left panel of Figure \ref{fig:potential} shows 
average potential magnetic field 
strength of X2.3 flare occurred on 2000 November 24.
This flux density is extrapolated by using 
the software package MAGPACK2 \citep{1982SoPh...76..301S}
and its strength is determined and averaged in the 
central portion of the flare.
In this calculation, the coronal flux density 
at the height of $ 2.56 \times  10^9 \  {\rm cm}$ that corresponds to 
the flare size of this event is $83\  {\rm G}$.
As shown in the left panel of Figure \ref{fig:potential}, 
the photospheric field 
strength is $335\  {\rm G}$ at $h=0$.
Note, however, that 
$\alpha_B$ in equation (\ref{eq:BBratio}) should 
be calculated by the ratio of the coronal flux density 
to the line-of-sight strength of the photospheric field.
Therefore we calculate the line-of-sight component of 
photospheric flux density using calculation of the potential field;
the spatial average of the line-of-sight component at $h=0$
in the flare site is $B_{\rm ph}=151\  {\rm G}$, which 
is shown as an asterisk in the left panel of Figure 
\ref{fig:potential}.
Accordingly, we obtain the ratio of 
coronal flux density to the line-of-sight 
component of photospheric flux density $\alpha_B$ is nearly 0.6.
In the right panel of Figure \ref{fig:potential},
we show the average potential magnetic field 
strength of M3.7 flare occurred on 2000 July 14 in the same way.
Concerning this event, the photospheric flux 
density is $214\  {\rm G}$ at $h=0$,
the coronal flux density 
at the height of the flare loop is $22\  {\rm G}$ and
the average line-of-sight component of photospheric field is
$B_{\rm ph}=119\  {\rm G}$;
$\alpha_B$ is roughly 0.2.
We calculate some other events and obtain the ratio
$\alpha_B$ in the range of  0.2 - 0.6.

Considering these estimates of the ratio, the assumption 
$\alpha_B \equiv B_{\rm cor}/B_{\rm ph} \sim 0.3 $ is 
probably plausible.
As a result, on the basis of the result described in \S
\ref{sec:result}, we find
the reconnection rate $M_{\rm A}$ is $10^{-3}$
for our analyzed events.

\subsection{Comparison with the Previous Study}
\label{sec:comparison}
In this section, we compare our results with 
the previous studies to evaluate our results. 
Using inflow pattern in extreme ultraviolet images,
the inflow velocity $V_{\rm in}$ was measured  
around $5 \  {\rm km \  s^{-1}}$ by \citet{2001ApJ...546L..69Y}, 
while \citet{Narukage_2006} obtained $2.6$ - $38$  ${\rm km \  s^{-1}}$ 
in six events.
Including other indirect measurements, 
the inflow velocity is in the same range as our results 
(Figure \ref{fig:m1_tL}).
Moreover, we analyze one of \citeauthor{Narukage_2006}'s events 
on 1999 May 27 using our method described in \S \ref{sec:howto}.
We estimate the inflow velocity at $5.2$ ${\rm km \  s^{-1}}$,
while they reported the inflow velocity of this event is 
$12.5$ - $14.9$ ${\rm km \  s^{-1}}$ or $31.3$ - $37.3$ ${\rm km \  s^{-1}}$. 
The former is obtained by the method based on \citet{2001ApJ...546L..69Y}, 
while the latter is obtained by the estimation based on 
\citet{2004ApJ...602L..61C}. 
In either case, the velocity obtained by our method is less than that obtained 
by their method.

Next, we compare our results with those of  \citet{2005ApJ...632..1184I},
since some of our analyzed events are also examined by them.
In our statistical study we use relatively simple methods 
to study a large number of events, 
while they dealt with three events in detail in their case study.
Table \ref{tab:compare} summarizes parameters of the flares 
obtained in this work and in \citet{2005ApJ...632..1184I},
mainly referring to their Table 1.
We calculate some parameters based on this table as follows.
\citet{2005ApJ...632..1184I} measured the width $L_x$, the
length $L_y$, and the height $L_z$ of the flare arcade separately.
To compare with our measurement of $L$, we calculate 
the flare size $L$ is defined as the cubic root of 
the product of $L_x$,$L_y$, and $L_z$ given in the paper.
The averaged energy release rate $E_{\rm flare}/\tau$ is 
defined as the thermal energy in the flare loop 
divided by the temporal scale,
\begin{equation}
E_{\rm flare}/\tau \equiv \frac{3}{2} n_{\rm loop} 
k_{\rm B}T L^3/\tau_{\rm flare} .
\end{equation} 
This value is calculated from
the peak value of $E_{\rm th}$ in 
\citeauthor{2005ApJ...632..1184I}'s 
Figures 6, 8, and 10 together with our {\itshape GOES} timescale 
$\tau_{\rm flare}$.
Moreover, we define the energy release rate $|dE_{\rm mag}/dt|$ 
as equation (\ref{eq:dEdt}) for our study and
as $H$ in \citet{2005ApJ...632..1184I} for their study,
both of which correspond to the Poynting flux into the 
diffusion region.
Furthermore it should be noted here that 
the {\itshape GOES} C-class event in Table \ref{tab:compare} is 
not one of our analyzed events in this statistical study 
due to lack of the non-saturated SXT image with the Be filter 
at the flare peak. Therefore using the last non-saturated image
with the Be filter taken at three minutes before the peak,
we calculate the parameters in this table for only comparison.

Comparing our result with \citet{2005ApJ...632..1184I},
we find that in our study the inflow velocity $V_{\rm in}$ 
is systematically smaller.
Since we calculate $V_{\rm in}$ as $L/(4 \tau_{\rm flare})$,
$V_{\rm in}$ obtained in our study is an average value in 
the impulsive phase of a flare.
On the other hand, \citet{2005ApJ...632..1184I} estimated
$V_{\rm in}$ using the peak value of the energy release rate
in the impulsive phase of the flares.
Therefore our small inflow velocity $V_{\rm in}$ perhaps reflects
nonsteadiness of the reconnection rate in the reconnection process.
Moreover, owing to our larger coronal magnetic flux 
density $B_{\rm cor}$,
the coronal Alfv\'en velocity $V_{\rm A}$ of our result
is systematically larger than those of their result.
Consequently, this larger $V_{\rm A}$ 
and smaller $V_{\rm in}$ make 
the reconnection rate $M_{\rm A}$ smaller than a factor of 2 - 10. 
Nevertheless, considering energy budget of each flare,
we note that our result is not inconsistent;
the energy release rate $|dE_{\rm mag}/dt|$ is comparable with 
the averaged energy release rate $E_{\rm flare}/\tau$
during the impulsive phase of the flare.

\subsection{Reconnection Rate and Reconnection Model}
\begin{figure}[htpb]
\plottwo{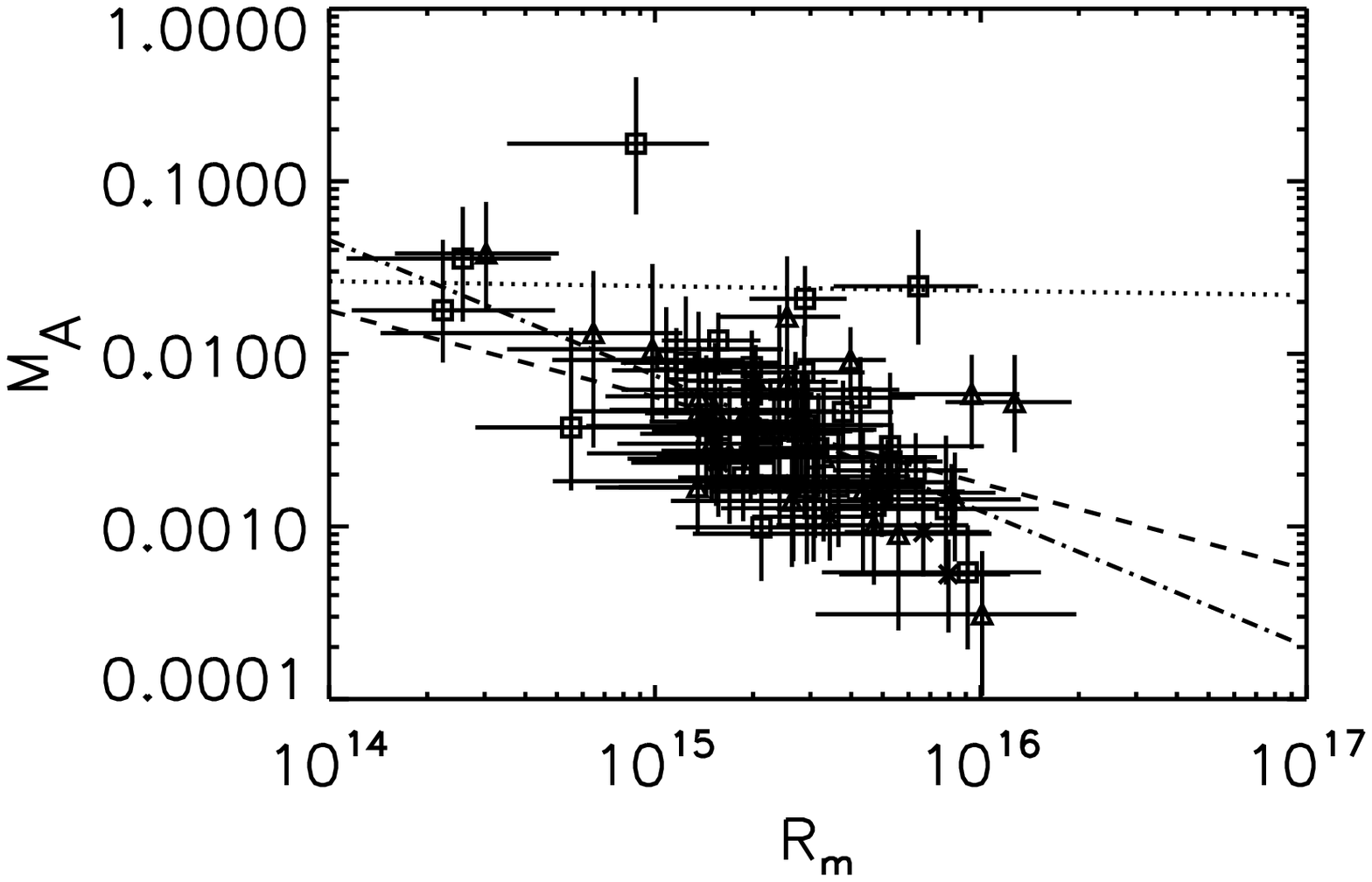}{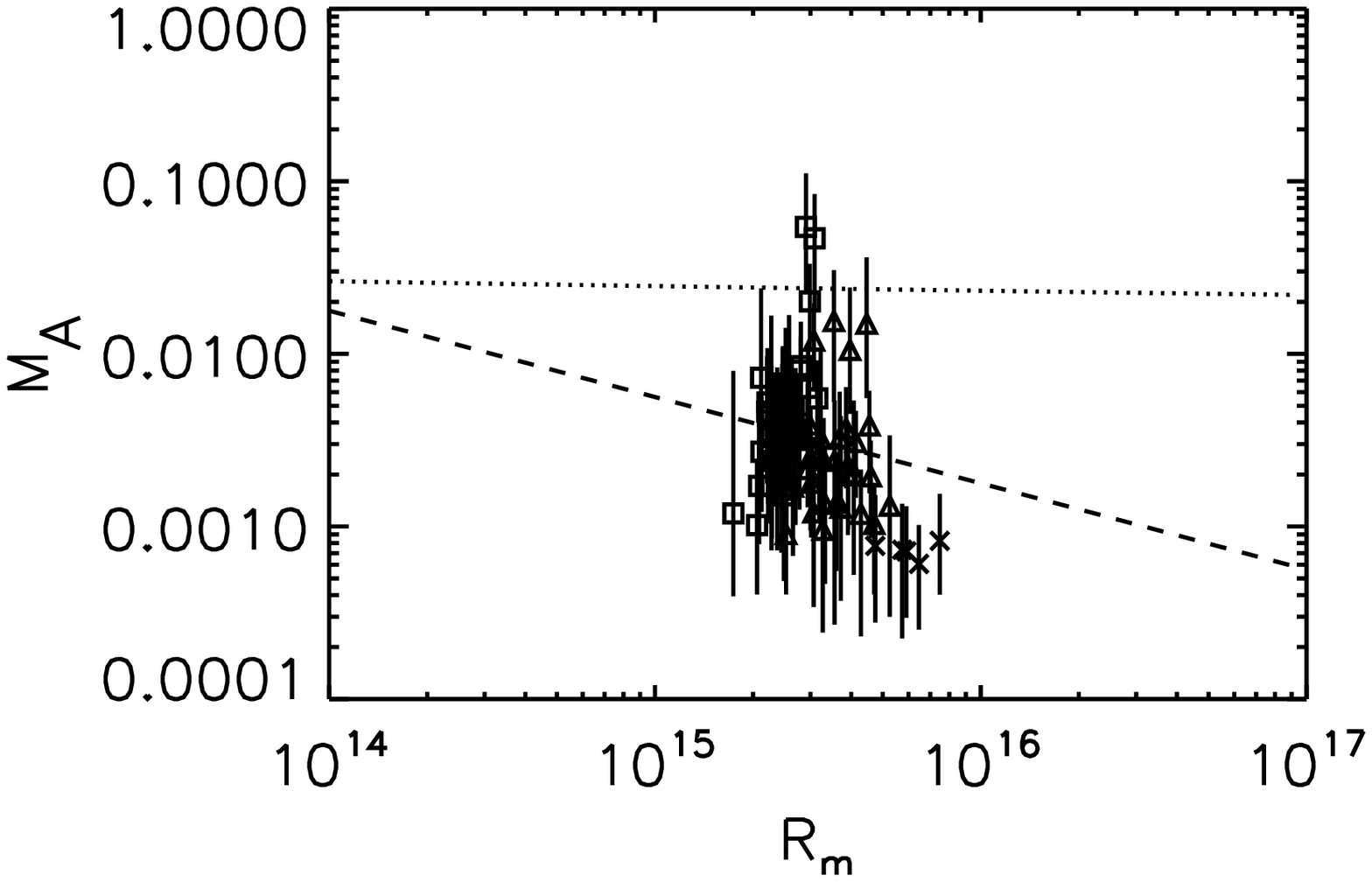}
\caption{Reconnection rate $M_{\rm A}$ plotted against the magnetic 
Reynolds number $R_m$. The left panel shows the reconnection 
rate obtained by method 1, and the right panel shows the 
reconnection rate obtained by method 2. In both panels, 
the dotted lines indicate the dependence of reconnection rate 
on the magnetic Reynolds number that is predicted by 
the Petschek model, and the dashed lines indicate 
the Sweet-Parker-type dependence $M_{\rm A} \propto  R_m^{-0.5}$.
The dash-dotted line in the left panel indicates the 
least-squares fitting of the data, which yields 
$M_{\rm A} \propto  R_m^{-0.8}$.
Meaning of the symbols is the same as Figure \ref{fig:m1_tL}.}
\label{fig:RmMa}
\end{figure}

Here we discuss the possible implication 
to the models from the reconnection rate 
obtained in this work. As mentioned in 
\S \ref{sec:intro}, different models 
show not only different values of the 
reconnection rate but also different 
dependence of the reconnection rate on the 
magnetic Reynolds number 
$R_{\rm m}=V_{\rm A}L/\eta$. 
Therefore we examine this dependence.
The reconnection rate $M_{\rm A}$ is plotted against the magnetic 
Reynolds number $R_{\rm m}$ in Figure \ref{fig:RmMa}.
The reconnection rate and the Reynolds number
obtained by method 1 is shown in the left 
panel of the figure and those obtained 
by method 2 is shown in the right panel. 
In the left panel, the dash-dotted line 
indicates the least-squares fitting of 
the data, which yields $M_{\rm A} \propto  R_{\rm m}^{-0.8}$,
while in the right panel there is no 
clear dependence since the magnetic Reynolds 
number is distributed within one order of magnitude.
To see the figure clearly, 
the error bars of the magnetic Reynolds number are omitted 
in the right panel, though the error is 
one order of magnitude at most.
The dotted lines in both panels indicate 
the dependence of the reconnection rate on the magnetic 
Reynolds number that is predicted by the 
Petschek model and the dashed lines indicate
the Sweet-Parker-type dependence 
$M_{\rm A} \propto  R_{\rm m}^{-0.5}$.
In calculation of the magnetic diffusivity 
$\eta$ in $R_{\rm m}$, we assume the 
resistivity is attributed to the Coulomb collision 
\citep{1956pfig.book.....S}
and the plasma temperature is $10^7 \  {\rm K}$.

In the process of estimating the reconnection rate $M_{\rm A}$ and 
the magnetic Reynolds number $R_{\rm m}$ shown 
in Figure \ref{fig:RmMa}, 
we have following assumptions:
(1) the coronal density $\rho$ is 
$1.67 \times 10^{-15} \  {\rm g \ cm^{-3}}$,
(2) the ratio of coronal flux density to photospheric flux density 
$\alpha_B \equiv B_{{\rm cor}}/B_{{\rm ph}}$ equals $0.3$, and
(3) the plasma temperature is $10^7 \  {\rm K}$.
If the coronal density $\rho$ and/or 
the ratio $\alpha_B$ deviate from the assumed value,
the values of the Alfv\'en velocity and 
the magnetic Reynolds number are affected.
However these values will only shift by some factors on the whole.
Therefore the distribution shown in Figure \ref{fig:RmMa}, i.e.,
the power index of the dependence of the reconnection rate 
$M_{\rm A}$ on
the magnetic Reynolds number $R_{\rm m}$ will 
not be greatly affected by assumptions (1) and (2).
Moreover, we can consider the factor to shift 
to be less than one order of magnitude,
because the coronal density $\rho$ is 
thought to be within at most one order of magnitude 
from the typical value and the ratio $\alpha_B$ is 
estimated in the range of 0.2 - 0.6 as mentioned in 
\S \ref{sec:alphadis}.
Furthermore, in analyzing by method 2,
we calculate the plasma temperature from {\itshape GOES} 
X-ray data and find that the observed 
temperature of plasma is distributed within 
less than one order of magnitude around $10^7 \  {\rm K}$. 
Hence assumption (3) is thought to have 
little effect on our results.

From the comparison of ours 
with the previous results mentioned in \S \ref{sec:comparison}
(e.g., \citealt{2005ApJ...632..1184I}; \citealt{Narukage_2006}), 
the error of the obtained reconnection rate 
shown in our figures is probably a little underestimated.
Besides the assumptions mentioned in the previous paragraph,
other uncertainties, such as the uncertainties 
in the measurements of the size $L$, the timescale 
$\tau_{\rm flare}$, the temperature $T$ and 
the emission measure $EM$, will affect our result.
Even after considering all these effects,
the value of $M_{\rm A}$ in Figure \ref{fig:RmMa} is found 
within one order of magnitude from the theoretical value 
of the Petschek model,
while dependence of the reconnection rate 
on the magnetic Reynolds number 
tends to be stronger than that in the Petschek model.
This result suggests two possibilities: 
One is the occurrence of the Petschek-type reconnection 
in the flares and 
the other is the Sweet-Parker model with the magnetic diffusivity
$\eta$ enlarged by some diffusion processes, such as 
the MHD turbulence.
Owing to the uncertainties, 
we cannot immediately rule out the possibility of 
either the Petschek model or the generalized Sweet-Parker model from our result.
We expect to discriminate between these two 
interpretations by the observations of {\itshape Solar-B}. 
If the Sweet-Parker reconnection with this 
enhanced resistivity occurs in a flare, 
the thickness of the diffusion 
region is estimated to be 100 km, 
when the length of the diffusion region is 
approximately the size of flare $L \sim 10^9 \  {\rm cm}$ 
and the reconnection rate is $M_{\rm A} \sim 0.01$. 
Because the size of diffusion region is less 
than the limit of resolution, the structure of turbulence 
is expected to be observed as a single-pixel line-broadening 
with {\itshape Solar-B}. On the other hand, 
if the Petschek-type reconnection occurs, the slow shock 
structure is expected to be observed.

\section{CONCLUSION} \label{sec:conclusion}
We analyze the data of flares occurred in the year of 2000 and 
measure the temporal scale $\tau$, the size $L$, and 
the magnetic field $B$ of each flare.
First, we examine the dependence of these 
parameters on the {\itshape GOES} class and find that 
the sizes of flares tend to be distributed more broadly
as the {\itshape GOES} class becomes weaker, and 
there is a threshold value in both photospheric and 
coronal magnetic flux density
that increases as the {\itshape GOES} peak flux increases.
Second, we examine the relationship between those parameters.
There is weak correlation between 
temporal and spatial scales of flares,
while there is little correlation between 
temporal scale and magnetic flux density.
Finally, we try to estimate the reconnection rate in each flare.
The values are distributed in the range from $10^{-3}$ to 
$10^{-2}$ and the reconnection rate decreases as the 
{\itshape GOES} class increases.
The value of the reconnection rate obtained in our study
is within one order of magnitude from 
the predicted maximum value of the Petschek model,
although the dependence of the reconnection 
rate on the magnetic Reynolds number tends 
to be stronger than that in the Petschek model.

\acknowledgments
We thank H. Isobe and T. J. Okamoto for careful reading and 
helpful comments.
We also thank K. Shibata for useful discussions.
We are grateful to T. Sakurai 
for allowing us to use his
potential field program (MAGPACK2).
Data analysis were carried out on the  
computer system at the Nobeyama Solar Radio Observatory of 
the National Astronomical Observatory of Japan and 
the Solar Data Archive (SODA) analysis server
at the PLAIN center, ISAS/JAXA.
The {\itshape Yohkoh} satellite is a Japanese national project, 
launched and operated by ISAS/JAXA, and 
involving many domestic institutions, with 
international collaboration with the US and the UK.
{\itshape Geostationary Operational Environmental Satellites}
({\itshape GOES}) 
is operated by the National Oceanic and Atmospheric 
Administration's (NOAA's) 
National Environmental Satellite, Data, and Information 
Service (NESDIS).
{\itshape SOHO} is a project operated by 
the European Space Agency and 
the US National Aeronautics and Space Administration.


\begin{thebibliography}{29}
\expandafter\ifx\csname natexlab\endcsname\relax\def\natexlab#1{#1}\fi

\bibitem[Chen et al.(2004)]{2004ApJ...602L..61C} Chen, P.~F., Shibata, K., 
Brooks, D.~H., \& Isobe, H.\ 2004, \apjl, 602, L61
 
\bibitem[{Dere(1996)}]{1996ApJ...472..864D}
Dere, K.~P. 1996, \apj, 472, 864

\bibitem[{{Domingo} {et~al.}(1995){Domingo}, {Fleck}, \&
  {Poland}}]{1995SoPh..162....1D}
{Domingo}, V., {Fleck}, B., \& {Poland}, A.~I. 1995, \solphys, 162, 1

\bibitem[{Garcia(1994)}]{1994SoPh..154..275G}
Garcia, H.~A. 1994, \solphys, 154, 275

\bibitem[Garcia(1998)]{1998ApJ...504.1051G} 
---. \ 1998, \apj, 504, 
1051
\bibitem[{{Isobe} {et~al.}(2002{\natexlab{a}}){Isobe}, {Morimoto}, {Eto},
  {Narukage}, \& {Shibata}}]{2002mwoc.conf..171I}
{Isobe}, H., {Morimoto}, T., {Eto}, S., {Narukage}, N., \& {Shibata}, K.
  2002{\natexlab{a}}, in Multi-Wavelength Observations of Coronal Structure and
  Dynamics, ed. P. C. H. Martens and D. P. Cauffman (Oxford: Pergamon), 171

\bibitem[{Isobe {et~al.}(2005)Isobe, Takasaki, \&
  Shibata}]{2005ApJ...632..1184I}
Isobe, H., Takasaki, H., \& Shibata, K. 2005, ApJ, 632, 1184

\bibitem[{{Isobe} {et~al.}(2002{\natexlab{b}}){Isobe}, {Yokoyama}, {Shimojo},
  {Morimoto}, {Kozu}, {Eto}, {Narukage}, \& {Shibata}}]{2002ApJ...566..528I}
{Isobe}, H., {Yokoyama}, T., {Shimojo}, M., {Morimoto}, T., {Kozu}, H., {Eto},
  S., {Narukage}, N., \& {Shibata}, K. 2002{\natexlab{b}}, \apj, 566, 528

\bibitem[Kay et al.(2003)]{2003A&A...400..779K} Kay, H.~R.~M., Harra, 
L.~K., Matthews, S.~A., Culhane, J.~L., \& Green, L.~M.\ 2003, \aap, 400, 
779

\bibitem[{Lin {et~al.}(2005)Lin, Ko, Sui, Raymond, Stenborg, Jiang, Zhao, \&
  Mancuso}]{2005ApJ...622..1251L}
Lin, J., Ko, Y.~K., Sui, L., Raymond, J.~C., Stenborg, G.~A., Jiang, Y., Zhao,
  S., \& Mancuso, S. 2005, \apj, 622, 1251

\bibitem[{{Matthaeus} \& {Lamkin}(1985)}]{1985PhFl...28..303M}
{Matthaeus}, W.~H., \& {Lamkin}, S.~L. 1985, Phys. Fluids, 28, 303

\bibitem[{Narukage \& Shibata(2006)}]{Narukage_2006}
Narukage, N., \& Shibata, K. 2006, \apj, 637, 1122

\bibitem[{Nitta(2004)}]{2004ApJ...610.1117N}
Nitta, S. 2004, \apj, 610, 1117

\bibitem[{{Ogawara} {et~al.}(1991){Ogawara}, {Takano}, {Kato}, {Kosugi},
  {Tsuneta}, {Watanabe}, {Kondo}, \& {Uchida}}]{1991SoPh..136....1O}
{Ogawara}, Y., {Takano}, T., {Kato}, T., {Kosugi}, T., {Tsuneta}, S.,
  {Watanabe}, T., {Kondo}, I., \& {Uchida}, Y. 1991, \solphys, 136, 1

\bibitem[{Ohyama \& Shibata(1997)}]{1997PASJ...49..249O}
Ohyama, M., \& Shibata, K. 1997, \pasj, 49, 249

\bibitem[{Ohyama \& Shibata(1998)}]{1998ApJ...499..934O}
---. 1998, \apj, 499, 934

\bibitem[{Parker(1957)}]{1957JGR....62..509P}
Parker, E.~N. 1957, \jgr, 62, 509

\bibitem[{Petschek(1964)}]{1964psf..conf..425P}
Petschek, H.~E. 1964, in The Physics of Solar Flares, ed. W.~N. Hess., 425

\bibitem[{Sakurai(1982)}]{1982SoPh...76..301S}
Sakurai, T. 1982, \solphys, 76, 301

\bibitem[{{Scherrer} {et~al.}(1995){Scherrer}, {Bogart}, {Bush}, {Hoeksema},
  {Kosovichev}, {Schou}, {Rosenberg}, {Springer}, {Tarbell}, {Title},
  {Wolfson}, {Zayer}, \& {MDI Engineering Team}}]{1995SoPh..162..129S}
{Scherrer}, P.~H., {Bogart}, R.~S., {Bush}, R.~I., {Hoeksema}, J.~T.,
  {Kosovichev}, A.~G., {Schou}, J., {Rosenberg}, W., {Springer}, L., {Tarbell},
  T.~D., {Title}, A., {Wolfson}, C.~J., {Zayer}, I., \& {MDI Engineering Team}.
  1995, \solphys, 162, 129

\bibitem[{{Shibata} \& {Tanuma}(2001)}]{2001EP&S...53..473S}
{Shibata}, K., \& {Tanuma}, S. 2001, Earth, Planets, and Space, 53, 473

\bibitem[{{Spitzer}(1956)}]{1956pfig.book.....S}
{Spitzer}, L. 1956, {Physics of Fully Ionized Gases} (New York: Interscience
  Publishers)

\bibitem[{Sweet(1958)}]{1958IAUS....6..123S}
Sweet, P.~A. 1958, in IAU Symposium no. 6, Electromagnetic Phenomena in
  Cosmical Physics, ed. B.~Lehnert (Cambridge University Press), 123

\bibitem[{{Tajima} \& {Shibata}(1997)}]{1997plas.conf.....T}
{Tajima}, T., \& {Shibata}, K. 1997, Plasma Astrophysics (Reading, Mass.
  :Addison- Wesley)

\bibitem[{Tsuneta(1996)}]{1996ApJ...456..840T}
Tsuneta, S. 1996, \apj, 456, 840

\bibitem[{{Tsuneta} {et~al.}(1991){Tsuneta}, {Acton}, {Bruner}, {Lemen},
  {Brown}, {Caravalho}, {Catura}, {Freeland}, {Jurcevich}, \&
  {Owens}}]{1991SoPh..136...37T}
{Tsuneta}, S., {Acton}, L., {Bruner}, M., {Lemen}, J., {Brown}, W.,
  {Caravalho}, R., {Catura}, R., {Freeland}, S., {Jurcevich}, B., \& {Owens},
  J. 1991, \solphys, 136, 37

\bibitem[{Tsuneta {et~al.}(1997)Tsuneta, Masuda, Kosugi, \&
  Sato}]{1997ApJ...478..787T}
Tsuneta, S., Masuda, S., Kosugi, T., \& Sato, J. 1997, \apj, 478, 787

\bibitem[{{Veronig} {et~al.}(2002){Veronig}, {Temmer}, {Hanslmeier}, {Otruba},
  \& {Messerotti}}]{2002A&A...382.1070V}
{Veronig}, A., {Temmer}, M., {Hanslmeier}, A., {Otruba}, W., \& {Messerotti},
  M. 2002, \aap, 382, 1070

\bibitem[{Yokoyama {et~al.}(2001)Yokoyama, Akita, Morimoto, Inoue, \&
  Newmark}]{2001ApJ...546L..69Y}
Yokoyama, T., Akita, K., Morimoto, T., Inoue, K., \& Newmark, J. 2001, \apj,
  546, L69

\end{thebibliography}


\clearpage
\begin{deluxetable}{cccccccccc}
\tabletypesize{\scriptsize}
\rotate
\tablecaption{Parameters of the Flares\label{tab:compare}}
\tablecolumns{10}
\tablewidth{0pt}
\tablehead{
 \colhead{Parameter} & \multicolumn{3}{c}{2000 Nov 24 15:13 X2.3 flare} &
\multicolumn{3}{c}{2000 Jul 14 13:52 M3.7 flare} &
\multicolumn{3}{c}{2000 Nov 16  0:40 C8.9 flare} \\
\colhead{}    &  
\multicolumn{2}{c}{This Work} &   
\colhead{\citet{2005ApJ...632..1184I}} &
\multicolumn{2}{c}{This Work} &   
\colhead{\citet{2005ApJ...632..1184I}}  &
\multicolumn{2}{c}{This Work} &   
\colhead{\citet{2005ApJ...632..1184I}}\\
\colhead{} & \colhead{Method 1}   &
 \colhead{Method 2}   & \colhead{} &
 \colhead{Method 1}   &
 \colhead{Method 2}   & \colhead{} &
 \colhead{Method 1}   &
 \colhead{Method 2}   & \colhead{}}
\startdata
$\tau$ (sec) & \multicolumn{2}{c}{1320} & & 
 \multicolumn{2}{c}{480}& &\multicolumn{2}{c}{1200}& \\
$B_{\rm ph}$(G) & \multicolumn{2}{c}{$205\pm 17$}  & 449 &
\multicolumn{2}{c}{$106\pm 11$} & 117 &
\multicolumn{2}{c}{$30 \pm 4 $} & 106 \\
$B_{\rm cor}$(G) & 62 & 116 & 41 & 32 & 60 & 44 &9 & 32 &11 \\
$L (10^9$cm) & \multicolumn{2}{c}{2.56 } & 2.8 & 
\multicolumn{2}{c}{2.94} & 2.3 &
 \multicolumn{2}{c}{4.12} & 4.0\\
$T (10^6$K) & \multicolumn{2}{c}{14} & 10 & \multicolumn{2}{c}{8.8}
 & 8.0 & \multicolumn{2}{c}{7.2} & 8.0\\
$V_{\rm in} {\rm (cm \ s^{-1})}$ &
\multicolumn{2}{c}{$4.8 \times 10^5$} &
$1.3 \times 10^7$ &\multicolumn{2}{c}{$1.5 \times 10^6$} &
$3.2 \times 10^6$ &\multicolumn{2}{c}{$8.6 \times 10^5$} &
$6.7 \times 10^6$\\ 
$V_{\rm A}{\rm (cm \ s^{-1})}$ & $4.25 \times 10^8$ & $8.0 \times 10^8$ & 
2.8 $\times 10^8$ &$2.2 \times 10^8$ & $4.2 \times 10^8$ & 
$2.1 \times 10^8$ & $6.2 \times 10^7$ & $2.3 \times 10^8$ &
$9.4 \times 10^7$\\
$M_{\rm A}$ & $1.1 \times 10^{-3}$ & $6.0 \times 10^{-4}$ &
 $4.7 \times 10^{-2}$ & $7.0 \times 10^{-3} $ &
$3.7 \times 10^{-3}$ & $1.5 \times 10^{-2}$ &
$1.4 \times 10^{-2}$ & $3.9 \times 10^{-3}$ & $7.1 \times 10^{-2}$\\   
$E_{\rm flare}/\tau {\rm (erg \ s^{-1}})$  &
 \multicolumn{2}{c}{$1 \times 10^{28}$} & $4 \times 10^{27}$&
 \multicolumn{2}{c}{$1 \times 10^{28}$} & $3 \times 10^{27}$&
 \multicolumn{2}{c}{$4 \times 10^{27}$} & $8 \times 10^{26}$ \\
$|dE_{\rm mag}/dt| {\rm (erg \ s^{-1}})$ & $2 \times 10^{27}$&
 $7 \times 10^{27}$&
$2.7 \times 10^{28}$\tablenotemark{a} & 
$2 \times 10^{27}$  & $8 \times 10^{27}$ &
 $6.4 \times 10^{27}$\tablenotemark{a}&
$2 \times 10^{26}$ & $2 \times 10^{27}$ &
 $1.9 \times 10^{27}$\tablenotemark{a} \\
\enddata
\tablenotetext{a}{Variable $H$ in \citet{2005ApJ...632..1184I}}
\end{deluxetable}
\clearpage

\end{document}